\documentclass{article}

\usepackage{graphicx}
\usepackage{fullpage,amsmath,amssymb,latexsym}
\usepackage{multirow}
\usepackage{natbib}

\begin{document}

\title{Jupiter's Moment of Inertia: A Possible Determination by JUNO}
\author{ Ravit Helled$^{1}$, John D. Anderson$^{2}$, Gerald Schubert$^3$, and David J. Stevenson$^{4}$\\
\small{$^1$Department of Geophysics and Planetary Sciences}\\
\small{Tel-Aviv University, Tel-Aviv 69978, Israel}\\
\small{$^2$Jet Propulsion Laboratory \footnote{Retiree},} \\
\small{California Institute of Technology, Pasadena, CA 91109}\\
\small{$^3$Department of Earth and Space Sciences}\\
\small{University of California, Los Angeles, CA 90095Ð1567, USA}\\
\small{$^4$ Division of Geological and Planetary Sciences} \\
 \small{California Institute of Technology, Pasadena, CA 91109}\\
\small{E-mail addresses: r.helled@gmail.com (R. Helled); jdandy@earthlink.net (J. D. Anderson)}\\
\small{schubert@ucla.edu (G. Schubert), djs@gps.caltech.edu (D. Stevenson)}\\
}

\date{}
\maketitle 

\begin{abstract}
The moment of inertia of a giant planet reveals important information about the planet's internal density structure and this information is not identical to that contained in the gravitational moments.    
The forthcoming Juno mission to Jupiter might determine Jupiter's normalized moment of inertia NMoI=C/MR$^2$ by measuring Jupiter's pole precession and the Lense-Thirring acceleration of the spacecraft ($C$ is the axial moment of inertia, and $M$ and $R$ are Jupiter's mass and mean radius, respectively). We investigate the possible range of NMoI values for Jupiter based on its measured gravitational field using a simple core/envelope model of the planet assuming that $J_2$ and $J_4$ are perfectly known and are equal to their measured values. The model suggests that for fixed values of $J_2$ and $J_4$ a range of NMOI values between 0.2629 and 0.2645 can be found. The Radau-Darwin relation gives a NMoI value that is larger than the model values by less than 1\%. 
A low NMoI of $\sim 0.236$, inferred from a dynamical model (Ward \& Canup, 2006, ApJ, 640, L91) is inconsistent with this range, but the range is model dependent.  Although we conclude that the NMoI is tightly constrained by the gravity coefficients, a measurement of Jupiter's NMoI to a few tenths of percent by Juno could provide an important constraint on Jupiter's internal structure. We carry out a simplified assessment of the error involved in Juno's possible determination of Jupiter's NMoI.
\end{abstract}


\newpage
\section{Introduction}
Jupiter's moment of inertia together with its mass, shape, size, and gravitational coefficients can provide key information about its current interior structure, formation, and evolution. 
A considerable amount of work has already been dedicated to the study of Jupiter's interior. 
Detailed interior models using theoretical hydrogen and helium equations of state (EOSs) have been used together with Jupiter's measured gravitational field ($J_2, J_4, J_6$) and other measured physical parameters (e.g., mass, radius, rotation rate) to better constrain its internal structure (e.g., Saumon \& Guillot, 2004; Guillot et al., 2004; Guillot, 2005; Fortney and Nettelmann, 2009; Militzer et al., 2008, Nettelmann et al., 2008). Yet, the heavy element enrichment of Jupiter's envelope and its core mass are not determined. \par

As more accurate EOSs and more complex models are developed, it is clear that more accurate gravitational data for Jupiter are desirable. 
An accurate determination of Jupiter's gravitational field (e.g., smaller error-bars and high order gravitational coefficients) could narrow the possible interior models and better constrain the planet's interior density profile. The forthcoming Juno orbiter  (Bolton, 2006) is planned to measure Jupiter's gravitational coefficients to high accuracy, and provide tighter constraints on the Jovian high-order gravitational coefficients.  
The coefficients do not directly constrain the density profile but rather the shapes of equipotential surfaces, which are in turn determined by the response of the planetary density to changes in pressure. This is indirectly a strong constraint on density but it is nonetheless possible to construct a model in which $J_2$ (for example) is left unchanged but the moment of inertia changes. So, while the density profile of a planet is often inferred from its measured gravitational coefficients $J_{2n}$, the axial moment of inertia $C$ (normalized moment of inertia NMoI=$C/MR^2$, where $M$ and $R$ are the planet's mass and mean radius, respectively) is an independent parameter that can be used to further constrain the density profile.   
The Juno spacecraft could determine Jupiter's NMoI by including the effects of the precession of Jupiter's pole and the Lense-Thirring acceleration on the Doppler signal measured by the X-Band and Ka-band systems. Below we investigate the possible range of NMoI values for Jupiter's measured gravitational field and its sensitivity to the core properties. Our conclusions are summarized in section 7. A simplified error analysis of Juno's determination of Jupiter's axial moment of inertia is presented in the appendix. 

\section{Juno's Measurements of Jupiter's NMoI}
Juno's radio science system (X-Band, Ka-Band) will provide an accurate determination of the Doppler shift during the 31 gravity-science orbits designed for a precise determination of Jupiter's gravitational field. 
Various effects influence the Doppler signal (see appendix); two of them involve Jupiter's NMoI. The first is the precession of Jupiter's pole and the second is the Lense-Thirring gravito-magnetism relativistic effect. In the following section we briefly summarize these two effects and their relation to Jupiter's NMoI.   

\subsection*{Location of Jupiter's Pole}
The precession of the spin pole of a rotating body like Jupiter can be determined by equating the rate of change of spin angular momentum with the applied gravitational torque. The equation of motion is given by $d{\bf s}/dt=\alpha({\bf n} \cdot {\bf s})({\bf s} \times {\bf n})$ where $\alpha$ is the precession rate, and ${\bf s}$ and ${\bf n}$ are the unit vectors along the spin pole and orbit normal, respectively. The precession rate varies with the spin angular momentum and the strength of the torque exerted by the Sun. For Jupiter, part of the solar torque is exerted on the Galilean satellites, which are locked close to Jupiter's equatorial plane. This "effective Jovian system" has a uniform precession rate  $\alpha$ (Goldreich, 1965) which is given by (e.g., Ward \& Hamilton, 2004),
\begin{equation}
\alpha = \frac{3}{2}\frac{GM_{\odot}}{\omega a_J^3} \left( \frac{J_2+Q}{\gamma+l} \right),
\end{equation}
where $G$ is the gravitational constant, $M_{\odot}$ is the Sun's mass, $\omega$ is Jupiter's rotation rate, $a_J$ is its semimajor axis, $J_2$ is the second gravitational harmonic coefficient, and $\gamma$ is Jupiter's normalized axial moment of inertia (NMoI). $Q$ is the gravitational contribution of Jupiter's satellite system to $J_2$, and $l$ is the angular momentum of the satellite system normalized to $MR^2\omega$ (see more details in Ward \&Canup, 2006 and references therein). 
Since Jupiter's orbital plane is not fixed in inertial space, its spin-axis precession changes with time.\par 
 
Figure 1 presents a simulation of the motion of Jupiter's pole due to torques from the Sun and Galilean satellites from 1975 to 2025 based on the IAU report for Jupiter's polar motion (Archinal  et al., 2011). Since the pole precession is dependent on Jupiter's NMoI its accurate determination by the Juno spacecraft during its 1-year lifetime can constrain Jupiter's moment of inertia. 
Figure 1 shows the simulated pole motion for two values of Jupiter's NMoI. The solid line represents NMoI of 0.25 while the dashed line uses a NMoI value decreased by 5\%.  Motion during the Juno orbiter phase is indicated by a thick solid line. 
Juno will be able to determine Jupiter's inertial polar location, along with the gravitational coefficients, during its gravity passes. Jupiter's polar motion can be determined by integration of Euler's equations (see Jacobson, 2002). The pole location is then determined by the integration constants for a given gravity orbit. The variation in the pole location for different orbits (i.e., times) is then used to determine Jupiter's precession rate. Jupiter's NMOI affects the equations of motion of the pole (see appendix for more details); 
it is therefore clear that an accurate measurement of Jupiter's pole motion (and therefore its precession rate) by Juno could put a useful constraint on Jupiter's moment of inertia and therefore, its internal mass distribution. The accuracy of this determination may depend on the a priori assumption that Jupiter has very low magnitude tesseral harmonics (i.e., very low deviation from rotational symmetry). \par

\subsection*{Lense-Thirring}  
The Lense-Thirring effect is a general relativistic (GR) effect which describes the amount of coordinate frame dragging by a rotating body (Lense \& Thirring, 1918). This effect is given by the linearization of the field equations of GR in which frame dragging is described in a form similar to that of Maxwellian electromagnetism. 
As a result, a point mass (Juno) gains Lense-Thirring acceleration $a_{LT}$ due to its close orbits of the massive rotating Jupiter.  
The Lense-Thirring acceleration is give by (Soffel, 1989), 
\begin{equation}
{\bf a}_{LT}= {\bf v} \times (\bigtriangledown \times {\bf \Phi})=\frac{K}{r^3}\left(
 \begin{array}{c}
    (x^2+y^2-2z^2/r^2)\dot y+3(yz/r^2)\dot z \\
    -(x^2+y^2-2z^2/r^2)\dot x-3(xz/r^2)\dot z  \\
    3(z/r)((x\dot y-y\dot x)/r)
 \end{array}
 \right)
\end{equation}
where ${\bf v}$ and ${\bf x}=(x,y,z)$ are Juno's velocity and position vectors, {\bf ($\dot{x},\dot{y},\dot{z}$)} are the time derivatives,  $r=|{\bf x}|$ is the radial distance, and ${\bf \Phi}\equiv-2G\frac{\bf J \times {\bf x}}{r^3}$. 
The constant K for GR is given by $K = 2\frac{GJ}{c^2}=\frac{1}{2}\frac{GM\omega R^2}{c^2}$
where $K$ is dependent on Jupiter's  rotational
angular momentum $J$, and $c$ is the speed of light. For a normalized moment of inertia of Jupiter equal to 1/4, Jupiter's angular momentum is given by $(1/4)MR^2\omega$, where for a
System III rotation period (the rotation period of Jupiter's magnetosphere; Jupiter's official rotation period) of 9hr 55m 29.71s, $\omega$ is 1.75853$\times10^{-4}$ s$^{-1}$. 
With $GM$ equal to 1.26687$\times10^8$km$^3$s$^{-2}$, mean radius $R$=69,940 km, and $c$ equal to 2.99792$\times10^5$km s$^{-1}$, the constant $K$ for Jupiter is
606.27 km$^3$ s$^{-1}$. \par

The Lense-Thirring effect accelerates the Juno spacecraft and it needs to be included in the Doppler signal analysis. 
The Lense-Thirring acceleration during gravity-science passes is a potentially measurable Doppler shift and could lead to a determination of Jupiter's NMoI (e.g., Iorio, 2010).  
A simplified Doppler error analysis which accounts for Jupiter's pole location and the acceleration due to Lense-Thirring is presented in the appendix. 

\section{The Radau-Darwin Approximation}
The NMoI values for planets are often estimated using the Radau-Darwin formula that relates the planetary NMoI and $J_2$, the second gravitational coefficient. Most giant planet modelers, however, do not use the Radau-Darwin approximation though it is occasionally used as a rough guide (e.g., Ward and Canup, 2006). 
The Radau-Darwin formula suggests that there is one-to-one correspondence between NMoI and $J_2$ by (e.g., Zharkov and Trubitsyn, 1978)  
\begin{equation}
NMoI= \frac{C}{MR^2}= \frac{2}{3}\left[1 - \frac{2}{5}\left( \frac{5m}{m+3J_2} -1 \right)^{1/2} \right].
\end{equation}
where $m=\omega^2R^3/GM$ is the small rotational parameter giving the ratio between centrifugal acceleration ($R$ is mean radius) and gravitational acceleration. 
For Jupiter, the Radau-Darwin relation suggests a NMoI of 0.2648, a value consistent with Jupiter having a small core (or no core at all). However, the Radau-Darwin relation is only a first order approximation. In fact, there is a range of NMoI values that can fit a given $J_2$. 

Ward \& Canup (2006) related Jupiter's obliquity to the precession of Uranus' orbit plane. Based on dynamical considerations, Jupiter's NMoI was inferred to be $\sim$0.236, considerably smaller than previous estimates and suggesting a massive core in Jupiter. 
It is therefore clear that an accurate determination of the possible NMoI values for Jupiter is desired. Below we use a simple core+envelope model of Jupiter constrained by Jupiter's gravity field ($J_2, J_4, J_6$) to explore the range of possible NMoI values. This exercise allows us to test the sensitivity of the NMoI to Jupiter's core properties (mass, radius), and in addition, allows us to test whether the NMoI value proposed by Ward \& Canup (2006) is feasible within the limitations of the model.

\section{NMoI Variation from A Simple Interior Model}
First, we construct an interior model of Jupiter. We use a simple model for Jupiter's density profile which consists of a constant density core and an envelope which is represented by a sixth order polynomial (e.g., Anderson \& Schubert, 2007). A discontinuity in the density function at the core-envelope boundary (CEB) is permitted. For a given core density and radius, the polynomial coefficients of the envelope density profile are found by iterating until the gravitational harmonics of the interior models converge to the measured ones (for details, see Helled et al., 2009). 
Table 2 summarizes Jupiter's physical parameters adopted in the model. \par 

The internal density is given by the normalized parameter $\delta = \rho/ \rho_0$ and the normalized mean radius is $\beta = s/R$, where $s$ is the mean radius internal to the planet. By using mean radius, the internal structure can be represented by the single parameter $\beta$, which labels the surfaces of constant total potential in the interior, the level surfaces (Zharkov and Trubitsyn, 1978). The total potential includes the internal gravitational potential plus the internal centrifugal potential. The core boundary is defined by the normalized radius $\beta_C$. The density in the envelope $\delta_E$ is given by a sixth degree polynomial,
\begin{equation}
\delta_E = a0 + a_2\beta^2 + a_3\beta^3 + a_4\beta^4 + a_5\beta^5 + a_6\beta^6.
\end{equation}\par

The first constraint on the interior model is that the density is zero at the surface. This is satisfied by setting the sum of the six polynomial coefficients to zero at $\beta=1$. The second constraint is that the integral of the density over the entire volume must equal Jupiter's mass. A third constraint is that the density derivative at $\beta=1$ equals the derivative of the density in the Lodders \& Fegley (1998) model atmosphere at the 1 bar pressure-level. For assumed core mass and radius, we then match the measured $J_2, J_4$ and $J_6$ as precisely as possible by using nonlinear least squares. The normalized axial moment of inertia for Jupiter can then be computed using the derived density distribution. The interior model does not account for differential rotation or shape deformation. Both of these effects add uncertainty to Jupiter's moment of inertia value and should be included in more advanced models. More details about the model can be found in Anderson \& Schubert (2007), Helled et al. (2009, 2011), and Schubert et al. (2011).\par

The model density profiles fit Jupiter's measured $J_2$ and $J_4$ {\it exactly} and fit its measured $J_6$ within its error bar. By assuming that $J_2$ and $J_4$ are perfectly known we can explore the variation of the NMoI value for a given gravitational field. Since Juno is predicted to provide a very precise determination of Jupiter's gravitational harmonics we can expect that in the future the variation in the allowed models for Jupiter will decrease due to tighter constraints on the measured gravitational field. We next search for a large range of Jupiter models that can match its measured gravitational field. Since the density profile is given by a mathematical function some of the models that satisfy Jupiter's physical parameters (i.e., mass, radius, gravitational filed) can be unphysical in terms of the density distribution. In order to exclude as many unphysical density distributions as possible we include the following constraints: (1) The density profile must be monotonic. (2) The core density $\rho_C$ cannot exceed 30 g cm$^{-3}$. (3) The core mass $M_C$ cannot exceed 40 M$_{\oplus}$. We then vary the core density and core radius and search for density profiles that satisfy the constraints listed above. 
The above constraints result in a maximum core radius of about 30\% of the planet's total radius. Our interior models are not based on physical equations of state that constrain the pressure-density relation (Saumon and Guillot, 2004; Nettelmann et al., 2008) and therefore our models consist of a broader range of values of core radius and mass.

\par

\subsection{Model Results}
Figure 2 shows the NMoI values that satisfy Jupiter's physical parameters and the additional constraints on the density distribution introduced above. The top of the figure shows the core mass in Earth masses ($M_{\oplus}$) vs. NMoI. The range of allowed models is shown in gray. Models with core density $\rho_C$ lower than the density at the core-envelope boundary $\rho_{CEB}$ are immediately excluded, together with models with core mass larger than the maximum core mass, and models with core density larger than the maximum core density . The bottom figure shows the range of NMoI for the allowed models as a function of core radius (normalized to Jupiter's total radius $R$). Not many models with small core radius can be found since the core density must stay lower than the maximum core density and yet be larger (or equal) than the envelope's density at the CEB. As the core radius increases more models can be found. However, in order to keep the core mass smaller than 40 M$_{\oplus}$ and at the same time keep the density profile monotonic, the number of models that can be found with increasing core radius decreases as the core radius exceeds $\sim$ 0.2$R$.  \par

We find that the NMoI value for Jupiter varies from 0.2629 to 0.2645, a 0.6\% variation. Figure 2 suggests that the ability to constrain Jupiter's core properties depends on the actual NMoI. If Jupiter's NMoI value is measured to be about 0.263-0.2635 our models suggest that Jupiter's core must be massive, on the order of 20 to 40 M$_{\oplus}$ with a core radius no smaller than 0.1 R. NMoI values between 0.2635 and 0.2640 can correspond to less massive cores with a larger possible range of core radii. If Jupiter's NMoI value is measured to be $\sim 0.2640$ our models suggest that Jupiter's core properties cannot be well constrained since core masses between zero and 40 M$_{\oplus}$ are possible, with radii that vary from about zero to 30\% of the planet's radius. Larger NMoI values correspond again to massive cores and large core radii but with a relatively low core density, on the order of 5-10 g cm$^{-3}$. On the other hand, NMoI values lower than $\sim 0.264$ correspond to relatively high core densities  (larger than 10 g cm$^{-3}$). If the density in the "core region" is not significantly larger than the density at the core-envelope boundary the model can be described as a "dissolved core model". The range of core masses for Jupiter based on recent interior models (Saumon and Guillot, 2004; Nettelmann et al., 2008; Milizer et al., 2008) ranges between 0 and 20 M$_{\oplus}$. For that range of core masses our model suggests a variation of $\sim$ 0.3\% for the NMoI value which is closer to the uncertainty  expected in the NMoI value measured by Juno (see appendix). This narrow range, however, is valid only for fixed values of the gravitational coefficients. Since Juno's measurement of Jupiter's gravitational field, in particular, $J_2$ and $J_4$, will include measurement uncertainties (i.e., error bars) the possible range of NMoI values will be larger. \par

Figure 3 shows the NMoI values in core mass- core radius parameter space. The colors correspond to Jupiter's NMoI values from the model. The solid blue curves correspond to constant density cores, i.e., curves of the form constant times r$^3$. The first curve from the left is for the maximum core density allowed,  30 g cm$^{-3}$. This curve is unique for a given maximum core density. The first curve from the right is defined by the constraint that the core density must be greater than or equal to the density at the CEB. This minimum density is actually a function of core radius and therefore this curve is non-unique; however, the minimum core density decreases only weakly with increasing core radius and the curve still provides an approximate constraint. The allowed models are constrained from above by the maximum allowed core mass. Also shown are curves for constant densities of 10 g cm$^{-3}$ and  20 g cm$^{-3}$ which typically correspond to ices and rocks, respectively. The results suggest that the NMoI value is about the same within the range of core masses between 0 and 10 M$_{\oplus}$ as predicted by models with physical equations of state (e.g., Saumon and Guillot, 2004). For this range of Jupiter's core mass a measurement of NMoI can hardly constrain Jupiter's core properties. More representative models of Jupiter are most likely to be between the two inner curves and for core masses between 0 and 20 M$_{\oplus}$. As the core mass increases the density-pressure relation in the envelope has to be adjusted to fit Jupiter's gravitational field. This adjustment to the polynomial coefficients are somewhat reflected by the figure's colors. For these cases the density-pressure relation in the envelope may not be representative for hydrogen/helium mixtures.


All the NMoI values we find for the allowed models are smaller than the Radau-Darwin value by less than 1\%. The lowest NMoI value we derive (0.2629, corresponding to a core mass of $\sim$ 40 M$_{\oplus}$) is still over 10\% larger than the value suggested by Ward \& Canup (2006). The NMoI value of Ward \& Canup (2006) is found to be inconsistent with physically plausible interior models of Jupiter that fit its measured $J_2$ and $J_4$ exactly. This result is in agreement with Cody \& Stevenson (2006). A measurement of Jupiter's NMoI by Juno can therefore   also provide constraints for dynamical models that relate Jupiter's spin-axis precession period and the precession period of Uranus' orbit plane (Ward \& Canup, 2006).  \par

All our models fit Jupiter's measured $J_6$ to within 3\%; this is about a five times smaller variation than the formal error bar on $J_6$. The models suggest a narrow range for Jupiter's $J_6$ value, between 34.69$\times10^{-6}$ and 35.71$\times10^{-6}$. However, this predicted range is based on the assumption of solid-body rotation, or that differential rotation does not not involve significant mass. Otherwise, the value of $J_6$ will include a dynamical contribution (Hubbard, 1999). Juno's measurement of Jupiter's gravitational field will test the validity of our prediction of the value of $J_6$. 

\section{Summary and Conclusions}

We use a simple core/envelope of Jupiter model to investigate the range of possible NMoI values that fit Jupiter's measured gravitational filed. Our results suggest  that a determination of Jupiter's NMoI with an accuracy on the order of a few tenths of percent could provide an additional constraint on Jupiter's internal density structure. The quality of the constraint will depend importantly on what the actual value of NMoI turns out to be. Our results suggest that Jupiter's NMoI is between 0.2629 and 0.2645, a range that depends on the model we use. Other models not encompassed within the boundaries of our model might allow for NMoI values outside our predicted range. The 0.6\% variation in our predicted range of allowed NMoI values is caused by different assumed core properties (i.e., mass or density). 
This range of NMoI values differs from the NMoI value obtained by the Radau-Darwin formula by less than $1\%$. Although the Radau-Darwin formula provides an excellent estimate of Jupiter's NMoI it unfortunately suggests that there is one-to-one correspondence between $J_2$ and NMoI. Our suite of models cannot accommodate the low value of 0.236 for Jupiter's NMoI suggested by Ward \& Canup (2006), but perhaps a different class of models can be made consistent with their value.\par

We point out that Jupiter's axial moment of inertia can be measured by the forthcoming Juno orbiter. Based on a simple error analysis that uses the 31 simulated gravity-science passes and their Doppler signals (see appendix for details), we estimate the standard error of the measured NMoI to be about 0.2\%. Our error analysis, however, is likely too optimistic, and the final error from the real data could be larger. 
Nevertheless if Jupiter's NMoI is indeed determined by the Juno spacecraft's Ka-Band transponder to an accuracy of $\sim$ 0.2\%, this measurement could provide an additional constraint on Jupiter's internal structure independent of the gravitational field measurements. Such a determination could provide valuable information on Jupiter's origin and evolution. 

\section{Appendix}
Below we present a simple analysis to estimate the error in Juno's measurement of Jupiter's NMoI. The analysis is a simple one and does not include all the variables and uncertainties that must be considered when Juno data become available. Juno's data analysis will need to include satellite-induced tides, high-order gravitational harmonics, tesseral harmonics, and other parameters/effects that are neglected in the analysis below. However, the error analysis below can be used as a first estimate for the accuracy of Jupiter's NMoI determination by Juno, and can provide a starting point for the more complicated Juno data analysis.

\subsection{Major Error Sources for a Measurement of the NMoI}
\label{ErrorSources}
There are two main sources of systematic error in the Doppler error analysis. The first is the error in the Juno orbit during the critical plus and minus three hours of closest approach, the perijove. The second is the error in the first three even zonal harmonics ($J_2,J_4,J_6$) in the gravitational field of Jupiter for altitudes of a few thousand kilometers above Jupiter's surface (defined as the 1 bar pressure level in the atmosphere (Lindal, 1992). These are not the only sources of error, but they are the most important because their neglect would lead to an overly optimistic assessment of the NMoI measurement. We neglect a number of other error sources for the reasons enumerated below. 

Systematic error enters into the radio Doppler data in the form of refraction by radio propagation through ionized and neutral media (Yakolev, 2002). The Juno Mission relies on an X-Band telecommunication system for spacecraft orbit determination and navigation, plus a Ka-Band system for gravity science. This configuration is similar to that used for radio science on the Cassini Mission to Saturn in its cruise phase between Earth and Saturn (Armstrong et al., 2003; Bertotti et al., 2003;  Anderson et al., 2004a, Anderson and Lau, 2010) before the Cassini spacecraft's Ka-Band system failed during cruise. Based on experience with that Cassini system, a conservative assumption for the Doppler error over a single perijove pass of six hours is 0.001 mm $\rm s^{-1}$ at a sample interval of 100 s, corresponding to an acceleration error of 10$^{-3}$ mgal. Another acceleration error estimate for a Doppler velocity measurement accuracy of $\sim$ 0.005 mm s$^{-1}$ at a sample interval of 60 s (Anderson et al., 2004b) is 10$^{-2}$ mgal. With the X-Band system alone, this error assumption must be increased by a factor of about 10-100. A good measurement of Jupiter's NMoI is impossible without the Ka-Band system.

Another source of error is uncertainties in the motion or ephemerides of the Earth and Jupiter in inertial space, and the motion of the tracking station fixed on the Earth's surface, with small but significant uncertainties in station location and in Earth's rotation and polar motion. A discussion of a number of small errors of this type are included in a report on the Pioneer anomaly (Anderson et al., 2002). The important consideration here is that we are interested only in errors that affect the Doppler signal over a period of six hours. To this end, methods for fitting radio Doppler data over short intervals of a few hours have been developed for the Doppler detection of gravitational radiation, in particular for the Ka-Band Cassini Ka-Band gravitational-wave experiment (Armstrong et al., 2003). Similar methods have been used for fitting short intervals of Doppler data for the Cassini flyby of Rhea (Anderson and Schubert, 2010). 
By applying some of these techniques to the Juno Doppler data, we expect to reduce the combined effect of all these small errors to a negligible level with respect to the assumed standard error of 0.001 mm $\rm s^{-1}$. The same can be said for  IR radiation from Jupiter and solar radiation perturbations to the motion of the spacecraft. 
These radiation sources produce a small but significant non-gravitational acceleration of the spacecraft . However, over a short interval of six hours this acceleration can be assumed constant and approximately in the Sun-Jupiter direction. An approximate acceleration model over six hours for the IR radiation from Jupiter, and a constant acceleration for the less troublesome solar radiation must be included in the fitting model, but the uncertainty in an assumed constant acceleration over six hours is negligible.

The last source of negligible error is the higher-order harmonics in the gravitational field of Jupiter. A complete gravitational field to high degree and order is needed for purposes of fitting the Doppler data to the noise level. In fact this is a major goal of Juno gravity science. In any error analysis for an orbiter or flyby of a giant planet, the important consideration is that the stimulation of gravitational harmonics under uniform planetary rotation is restricted to the even zonal harmonics in a general expansion of the gravitational potential V according to Kaula (1968),   
\begin{equation}
 V = \frac{G M}{r} \left( 1+ \sum_{\ell = 2}^\infty \sum_{m = 0}^\ell \left( \frac{a}{r} \right)^\ell P_{\ell m} \left( \sin \phi \right) \left[C_{\ell m} \cos m \lambda + S_{\ell m} \sin m \lambda \right] \right)
\label{V}
\end{equation}
where $G$ is the gravitational constant, $M$ is the total mass of the planet ($ C_{0 0} = 1$), $C_{\ell m}$ and $ S_{\ell m}$ are constants of integration for Laplace$^\prime$s equation, $ P_{\ell m}$ is the associated Legendre polynomial, and $a$ is a reference radius for the planet such that $C_{\ell m}$ and $ S_{\ell m}$ are dimensionless. It is customary to take the origin of coordinates at the center of mass so that the first-degree integration constants are zero by definition, and $a$ is the equatorial radius a of the planet. The spherical coordinates of an external point in the potential are the radius $r$, the latitude $\phi$ and the longitude $\lambda$, all three referred to the center of mass. 
Our simplified approach assumes that equation (5) can be approximated by a truncated series involving only the even zonal harmonics,  
\begin{equation}
V =  \frac{G M}{r} \left[ 1 - \sum_{i = 1}^n J_{2 i} \left( \frac{a}{r} \right)^{2 i} P_{2 i} \left( \sin \phi \right)   \right]
\label{Vrot}
\end{equation}
The truncation number n is taken sufficiently large that the rotational stimulation of harmonics larger than 2n is negligible, with respect to the error in the observations. 

For the simplified error analysis we consider only the first three coefficients $J_2$, $J_4$ and $J_6$. All other constants in Eq.~5 are assumed to contribute negligible error, with the following justification. The product $GM$ for Jupiter is known currently to a fractional error of better than $2 \times 10^{-8}$ (Jacobson, 2003, JUP230 orbit solution, $\rm$ http://ssd.jpl.nasa.gov/?gravity\_fields\_op), and it will be improved substantially with the inclusion of Juno Doppler data in a general solution for masses in the Jupiter system. The zonal coefficients other than the first three, and all the other integration constants of Eq.~5.,  contribute Doppler noise at Fourier frequencies much higher than $J_2$, $J_4$, the spacecraft orbit, Jupiter pole location, and the NMoI, which are all included in  our error analysis. We include J$_6$ in the error analysis mainly to prove that it is uncorrelated from the other included parameters, and that harmonics of degree 6 and higher contribute no significant error to the determination of the NMoI. The power spectral density at low Fourier frequencies for these other harmonics is smaller than the Doppler noise, assumed white. This does not mean the other harmonics are insignificant. The odd zonal harmonics can be stimulated by deep zonal flows in the interior of Jupiter, and internal dynamics in general can stimulate all harmonics (Kaspi et al., 2010;  Hubbard, 1999). An objective of the Juno gravity experiment is to characterize the contribution of this high-frequency noise in Jupiter's gravitational field, expected to have a standard error in acceleration g on the order of one mgal at the one-bar level, and to gain information on internal dynamics. However, this goal is independent of the determination of the NMoI and is not considered here. 

\subsection*{Methodology of the Error Analysis}
We use a current set of perijove orbits delivered to the Juno Science Team by JPL. These are simulated orbits, of course, but they can be used to assess the possibility of an NMoI measurement during the projected science phase of the mission from 10 November 2016 to 5 October 2017. There are 31 perijove opportunities in all, separated by about 10.97 d. Ka-Band Doppler tracking is anticipated on many of these perijove passages, but not all. For passes dedicated to microwave radiometry, the spacecraft high-gain antenna must point at Jupiter, not Earth, and Doppler tracking is impossible. Nevertheless, we anticipate 22 or so good Doppler passes over the duration of the science phase. At each perijove, simulated orbital elements are available as classical inclination I, node $\Omega$, and perijove location $\omega$ (Danby, 1992) with respect to the true Jupiter equator and equinox of date. The orbits are polar with inclinations between 90 and 91$^o$. The latitude of the closest approach point is 5.7$^o$ at the start of the science phase and it precesses north to a latitude of 34.2$^o$ at the end. The orbital motion during perijove passage is from north to south. The direction of the Earth at perijove can be found from planetary ephemerides of Earth and Jupiter. The Earth phase angle with respect to the orbit normal varies from about 17$^o$ early in the science phase to about 30$^o$ at its end. The orbit is not viewed in its orbit plane, hence the effects of the zonal harmonics on the Doppler data are reduced, but effects that vary with Jupiter longitude are increased, such as the polar motion and Lense-Thirring rotational frame dragging. Given this trade off, the Earth angle is near optimal for gravity science.   

In addition to the orbital elements at perijove, the radial distance r is also given  and varies from 75,102 km near the beginning of the science phase to 76139 km at its end. The corresponding altitudes above the Jupiter geoid are 3748 km and 6214 km, respectively. We convert these orbital elements into inertial position and velocity six days prior to perijove, or three days before the start of Doppler tracking for gravity science. The magnitude of the velocity vector is not given, but it can be adjusted so that the orbital period for our orbital model is consistent with a period of about 11 d. Our orbital model consists of Cartesian accelerations by means of the gradient of the potential with the current values of $GM$, $J_2$, $J_4$, $J_6$ included (Helled et al., 2009), as well as the Lense-Thirring acceleration for an NMoI of 0.24. The Lense-Thirring equations of motion are obtained by a linearization of the Kerr metric and a solution of the geodesic equations, and they agree with equations given by Soffel (1989), although Soffel uses an NMoI of 2/5, which is too large for Jupiter. The conversion of his proportionality constant K to an NMoI of 0.24 involves nothing more than a simple scaling by 0.24/0.4. The precession of the pole is taken into account in the equations of motion as well. We assume the direction of the polar motion is known and include two unknown constants $u_0$ and $\dot{u}$ for a linear drift $u_0 + \dot{u} \left( t - t_0 \right)$ in the direction given by the IAU pole precession in right ascension and declination, which we convert to a direction for the IAU Jupiter equator and equinox (Seidelmann et al., 2002). The epoch $t_0$ is taken at the closest approach time for the mid-mission perijove pass on 13 April 2017. The angle to Jupiter's space-fixed $X$ axis is 139$^o$ and the precession rate is 64.5 $ \mu rad~cy^{-1}$. In the real data analysis, the precession can be modeled more accurately by solving Euler's equations with torques from the Sun and satellites. However, for an error analysis the standard IAU linear model is adequate (see 2009 IAU/IAG report for the Jupiter polar motion, and Jacobson, 2002 for details). An assumption that the precession direction is unknown and must be included in the error analysis leads to an almost perfect correlation (0.9999) between NMoI and pole location on each perijove pass. The high correlation is surprising, but perhaps not too unexpected. Both pole location and Lense-Thirring precession are inertial effects on the Doppler signal. The precession angle is not known exactly, of course, but a 10$^o$ variation about its assumed direction of 139$^o$ does not affect the Doppler signal significantly. A 10$^o$ variation is a factor of 10 or more less than the effect of other parameters included in the error analysis.

The simulated Doppler signal, sampled at an interval of 100 s, is shown in Fig.~4 for the perijove pass on 21 November 2016, the first gravity pass of the science phase. Other perijove passes yield a similar Doppler signal. With a Ka-Band error of 0.001 mm s$^{-1}$, it is possible to detect fractional perturbations to this Doppler signal on the order of 10$^{-9}$. This is good, but it could also present problems in that there might be systematic error in the signal that is very difficult to understand. The Earth flyby anomaly comes to mind, which was first detected in 1990 and is still not understood (Anderson et al., 2007). 

The error analysis proceeds by standard methods (Tapley, 1973). We vary the parameters for the error analysis one at a time and compute the perturbation to the Doppler curve at the sampled points. The results are put in a design matrix $A$, column by column for each parameter. The matrix $A$ is then decomposed by singular value decomposition (SVD) such that $A = U.W.V ^T$ (Hestenes, 1958), where the superscript $T$ indicates a transpose . In the analysis of real data, the decomposed rectangular matrix $A$ can be inverted and multiplied by the residual matrix $z$ to obtain corrections x to the current guess at the parameters according to $x = A^{-1}.z$  (Lawson, 1974). The inverse of $A$ is called the pseudo inverse. In an error analysis, the covariance matrix $\Gamma$ can be computed by the inversion of a diagonal matrix representing the singular values of $A$. The result for a uniform weighting of the range-rate observations is,
\begin{equation}
\Gamma = \sigma_{\dot{\rho}}^2 V \left( W^T W \right)^{-1} V^T
\label{CovMat}
\end{equation}
In the data analysis the design matrix is recomputed at each iteration on the parameters until everything converges, in the sense that the residuals are significantly smaller than the errors on the data, and the corrections $x$ are sufficiently small. For the error analysis it is assumed that the design matrix has converged and that $\Gamma$ represents a converged covariance matrix.

The low-frequency parameters in the error analysis are the six components of the Cartesian position and velocity, taken six hours before closest approach, the gravity coefficients $J_2$, $J_4$ and $J_6$, the pole location $u_0$, and the NMoI which affects the rate of both the polar precession and the Lens-Thirring precession, which is directly proportional to the NMoI. The polar precession is inversely proportional to the NMoI. A spheroid of constant density precesses more slowly than a spheroid that is centrally condensed. There are a total of 11 parameters in the covariance matrix for each six-hour perijove pass. The numerical properties of the design matrix $A$ are conditioned by using units on each parameter such that the Doppler curve is perturbed on the order of a few mm s$^{-1}$. The spacecraft initial position is perturbed by 10 m in all three Cartesian components, and similarly the velocity components are perturbed by 1 mm s$^{-1}$. The three zonal gravity coefficients are perturbed by 10$^{-6}$. The pole location is perturbed by 10 $\mu$rad and the NMoI by 1000\%. A 1000\% perturbation in the NMoI is required in order to obtain Doppler variations comparable to those of the other parameters in the error analysis. Comparable variations are desirable in order to guard against numerical degeneracies in the covariance matrix, as opposed to physical degeneracies. The partial derivatives for these last two parameters are shown in Fig.~5 and Fig.~6. The low-frequency nature of these two parameters is evident. For $J_6$ and even for $J_4$, the high-frequency components are not easily resolved with a 100 s sample interval. 
Juno's Radio Science Ka-Band data are expected to be delivered at a sample interval of one second. For our investigation 10 s intervals are used. However, sample intervals larger than one second can be used by averaging the data over larger intervals. Filtering the data by a moving point average is similar, except that resulting smoothed data at a sample interval of one second are correlated. The data are independent when averaged in intervals instead. 

However, all the parameters in the error analysis exhibit the low-frequency nature of Fig.~5 and Fig.~6. The singular values of the matrix $A$ for the 21 November case represented by the figures are 128.55, 26.4578, 11.4494, 10.9035, 7.09225, 5.96269, 4.11035, 
3.18769, 1.25676, 0.102384, and 0.0000205466. It is this last singular value that is of concern. There is a combination of the parameters for which the data offer very little information. In fact we expect this behavior from an orbiter. For a spherical planet the orbit can be rotated about the line of sight and there will be no change in the Doppler signal. This is comparable to the fact that the radial velocity of a spectroscopic binary star is independent of its node on the plane of the sky (Hilditch, 2001). For orbiters of planets, the degeneracy is broken partially by the spheroidal shape of the planet, and by the motion of the planet on the sky, but in this Juno error analysis both the oblateness of the planet and the orientation of the pole are uncertain. There is indeed one degree of degeneracy. Rather than discard this one singular value, and perhaps obtain a covariance matrix that is too optimistic, we instead replace the smallest singular value by 1.0 in the matrix $W^T W$, comparable to the ninth singular value. The idea here is that the Juno navigation team has some information on the orbit at perijove from Doppler tracking outside the plus and minus three hours of gravity data. We can use their best estimate of the node on the plane of the sky as a-priori information on the orbit. However, the lack of information on this one singular value does not affect the error estimate on the constants of interest, but only the orbit. The proof is contained in the eigenvector associated with the minimum singular value. That vector is $0.42672 \Delta X_0 - 0.31053 \Delta Y_0 + 0.78199  \Delta Z_0 -0.03565 \Delta \dot{X}_0 + 0.03949 \Delta \dot{Y}_0 - 0.32731  \Delta \dot{Z}_0 - 0.00173  \Delta u_0 - 0.00142  \Delta NMoI$. The effect on the pole location and the NMOI is small, and the effect on the zonal harmonics is essentially zero. It is reasonable to increase the orbital information by increasing the information on the smallest singular value.

The covariance matrix of Eq.~7 can be evaluated for all the perijove passes with gravity Doppler tracking. The procedure we use to combine the information from a number of passes is to first extract the covariance matrix for the five constants. The orbital initial conditions are different on each perijove pass, and midcourse maneuvers and spacecraft activity by the mission team destroys any possibility of establishing coherency between perijove orbits, for example to look for long-term orbital precession terms over one year. The idea here is that systematic error from the orbit uncertainties are contained in the $11 \times 11~\Gamma$ matrix. All that is required is the $5 \times 5$ sub matrix of $\Gamma$ that represents the five constants $J_2$, $J_4$ $J_6$, $u_0$ and NMoI. An inversion of that sub matrix contains all the Doppler information from a single pass and also contains systematic error from uncertainties in the orbit. These inverted matrices can be added together, and then their sum can be inverted for purposes of obtaining the covariance matrix on the five constants from several passes. The anticipated data analysis proceeds in much the same way, with both the orbit and the constants determined on each perijove pass, and then just the constants statistically combined in order to obtain a result for the overall Juno science mission.
Figure 6 shows the standard error in NMoI as the gravity-science orbits are combined. It can be seen from the figure that combining the data can improve the standard error by a factor of $\sim$ 2.5. If data from all the planned gravity-science passes are collected, we suggest that Jupiter's NMoI could be measured to an accuracy of 0.22\%. 

\section*{References}
Anderson, J. D. , Campbell, J. K. and Nieto, M. M., 2007. The energy transfer process in planetary ßybys. 
New Astronomy, 12, 383--397.\\ 
Anderson, J. D., Laing, P. A. , Lau, E. L., Liu,  A. S., Nieto, M. M., and Turyshev, S. G., 2002. Study of the 
anomalous acceleration of Pioneer 10 and 11. Phys. Rev. D, 65(8), 082004.\\ 
Anderson, J. D.  and Lau, E. L., 2010. Measurements of Space Curvature by Solar Mass. In I. Ciufolini \& 
R. A. Matzner, editor, General Relativity and John Archibald Wheeler, volume 367 of Astrophysics and 
Space Science Library, 95--108, 2010. \\
Anderson, J. D., Lau, E. L., and Giampieri, G., 2004a. Measurement of the PPN parameter with radio signals 
from the cassini spacecraft at X and Ka bands. In P. Chen, E. Bloom, G. Madejski, and V. Petrosian, 
editors, 22nd Texas Symposium on Relativistic Astrophysics, Menlo Park, CA, 2004. Stanford Linear 
Accelerator Center Technical Publications. \\
Anderson, J. D., Lau, E. L., Schubert, G., Palguta, J. L., 2004b. Gravity Inversion Considerations for Radio Doppler Data from the JUNO Jupiter Polar Orbiter. 
American Astronomical Society, DPS meeting 36, 14.09; Bulletin of the American Astronomical Society, Vol. 36, p.1094.\\
Anderson, J. D., and Schubert, G. 2007, Saturn's gravitational field, internal rotation, and interior structure. Science, 317, 1384--1387. \\
Anderson, J. D. and Schubert, G., 2010. Rhea's gravitational Þeld and interior structure inferred from archival 
data Þles of the 2005 Cassini ßyby. Physics of the Earth and Planetary Interiors, 178, 176--182.\\ 
Archinal, B. A., A'Hearn, M. F., Bowell, E., Conrad, A., Consolmagno, G. J., Courtin, R., Fukushima, T., Hestroffer, D., Hilton, J. L., Krasinsky, G. A., Neumann, G., Oberst, J., Seidelmann, P. K., Stooke, P., Tholen, D. J., Thomas, P. C., and Williams, I. P., 2011. Report of the IAU Working Group on Cartographic Coordinates and Rotational Elements: 2009. 
Celestial Mechanics and Dynamical Astronomy, 109, 101--135.\\
Armstrong, J. W., Iess, L., Tortora, P. and Bertotti, B, 2003. Stochastic Gravitational Wave Background: 
Upper Limits in the 10$^{-6}$ to 10$^{-3}$ Hz Band.  The Astrophysical Journal, 599, 806--813.\\
Bertotti, B., Iess, L. and Tortora, P., 2003. A test of general relativity using radio links with the cassini 
spacecraft. Nature, 425, 374--376.\\
Bolton, S. (2006). The Juno New Frontiers Jupiter polar orbiter mission. European Planetary 
Science Congress, 535. \\
Cody, A. M. and Stevenson, D. J., 2006. Constraints on the Moment of Inertia of Jupiter via Interior Models. American Astronomical Society, DPS meeting \#38, \#11.07; 
Bulletin of the American Astronomical Society, Vol. 38, p. 496\\
Danby, J. M. A., 1992. Fundamentals of celestial mechanics. Richmond: Willman-Bell, 2nd ed., 1992.\\
Guillot, T., 2005. The interiors of giant planets: Models and outstanding questions. Annu. Rev. 
Earth Planet. Sci., 33, 493--530. \\
Guillot, T., Stevenson, D. J., Hubbard, W. B. and Saumon, D., 2004. The interior of Jupiter. In 
Jupiter: The Planet, Satel lites and Magnetosphere, eds. F. Bagenal, T. E. Dowling and W. B. 
McKinnon, Cambridge Planetary Science, Cambridge, UK: Cambridge University Press,  
35--57.\\ 
Helled, R., Schubert, G. and Anderson, J. D., 2009. Empirical models of pressure and density in 
SaturnÕs interior: Implications for the helium concentration, its depth dependence, and SaturnÕs 
precession rate. Icarus, 199, 368--377. \\
Helled, R., Anderson, J. D., Podolak, M., and Schubert, G., 2011.Interior Models of Uranus and Neptune.  The Astrophysical Journal, 726, 15--16.\\
Hestenes, M. R., 1958. Inversion of matrices by biorthogonalization and related results. Jour. SIAM, 6, 51--90.\\ 
Hilditch, R. W. 2001 An Introduction to Close Binary Stars. Cambridge University Press, Cambridge, 
UK, 2001. \\
Hubbard,  W. B, 1999. NOTE: Gravitational Signature of JupiterÕs Deep Zonal Flows. Icarus, 137, 357--359.\\
Iorio, L., 2010. Juno, the angular momentum of Jupiter and the Lense-Thirring effect. New Astronomy, 15, 544--560.\\
Jacobson, R. A., 2001. The Gravity Field of the Jovian System and the Orbits of the Regular Jovian Satellites. 
In Bul letin of the American Astronomical Society, volume 33 of Bul letin of the American Astronomical 
Society, 1039. \\
Jacobson, R. A., 2002. The Orientation of the Pole of Jupiter.  American Astronomical Society, DDA Meeting 33,  07.02; 
Bulletin of the American Astronomical Society, Vol. 34, p.936.\\
Jacobson, R. A. 2007, in AAS/Division of Dynamical Astronomy Meeting, Vol. 38, AAS/Division of Dynamical Astronomy Meeting, 13 \\
Kaspi, Y., Hubbard, W. B. , Showman, A. P. and Flierl, G. R., 2010. Gravitational signature of JupiterÕs internal 
dynamics. Geophys. Res. Lett., 37, 1204.\\ 
Kaula, W. M. 1968, An introduction to planetary physics - The terrestrial planets (Space Science Text Series, 
New York: Wiley, 1968) \\
Lawson, C. L.  and Hanson, R. J., 1974.  Solving least squares problems. Prentice-Hall, 1974. \\
Lense, J. \& Thirring, H., 1918. Phys. Z. 19, 156. Translated and discussed in Mashhoon, 
B., Hehl, F.W., Theiss, D.S., 1984. Gen. Relativ. Gravit. 16, 711. Reprinted in 
RufÞni, R.J., Sigismondi, C. (Eds.), 2003. Nonlinear Gravitodynamics. World 
ScientiÞc, Singapore, 349--88. \\
Lindal, G. F. 1992, The atmosphere of Neptune - an analysis of radio occultation data acquired with Voyager 
2.  The Astrophysical Journal, 103, 967--982.\\
Lodders, K., \& Fegley, B. 1998, The planetary scientistÕs companion. New York : Oxford University 
Press, 1998. QB601 .L84 1998) \\
Militzer, B., Hubbard, W. B., Vorberger, J., Tamblyn, I. and Bonev, S. A., 2008. A massive core 
in Jupiter predicted from Þrst-principles simulations.  The Astrophysical Journal, 688, L45--L48.\\
Nettelmann, N., Holst, B., Kietzmann, A., French, M. and Redmer, R., 2008. Ab inition equation 
of state data for hydrogen, helium, and water and the internal structure of Jupiter.  The Astrophysical Journal, 683, 1217--1228. \\
Saumon, D. and Guillot, T. , 2004. Shock compression of deuterium and the interiors of Jupiter 
and Saturn.  The Astrophysical Journal, 609, 1170--1180.\\ 
Schubert, G., Anderson, J. D., Zhang, K., Kong, D., Helled, R., 2011. 
Shapes and gravitational fields of rotating two-layer Maclaurin ellipsolids: Application to planets and satellites. Physics of the Earth and Planetary Interiors, in press.\\
Soffel, Michael H., 1989. Relativity in Astrometry, Celestial Mechanics and Geodesy. Astronomy and Astro- 
physics Library. Springer-Verlag, Berlin, 1989. \\
Tapley, B. D., 1973. Statistical Orbit Determination Theory. In B. D. Tapley \& V. Szebehely, editor, Recent 
Advances in Dynamical Astronomy, volume 39 of Astrophysics and Space Science Library, 396,  
1973. \\
Ward, W. R., \& Hamilton, D. P. 2004. Tilting Saturn. I. Analytic Model. The Astrophysical Journal, 128,  2501--2509. \\
Ward \& Canup, 2006. The Obliquity of Jupiter. The Astrophysical Journal, 640,  L91--L94. \\
Yakolev, O. I., 2002. Space Radio Science. Earth Space Institute Book Series. Taylor \& Francis, London, 
2002. Translated from the Russian by Nikolai and Olga Golovchenko. \\
Zharkov, V. N., \& Trubitsyn, V. P. 1978, Physics of planetary interiors (Astronomy and Astrophysics Series,
Tucson: Pachart, 1978)

\newpage
\begin{table}
\begin{center}
\begin{tabular}{lccl}
\multicolumn{2}{c}{} \\
\hline
\hline
Parameter & Jupiter at 1 bar\\
\hline
GM (km$^3$ s$^{-2}$) & 126,712,765 $\pm2$\\
r$_e$ (km)  & 71,492  $\pm4$\\
r$_p$ (km) & 66,854 $\pm4$ \\
J$_2$ (10$^{-6}$) &14696.43 $\pm0.21$\\
J$_4$ (10$^{-6}$) & -587.14 $\pm1.68$\\
J$_6$ (10$^{-6}$) & 34.25 $\pm5.22$\\
\hline
\end{tabular}
\caption[Hello]{
	\label{S}
Jupiter's data as given by JPL database:  http://ssd.jpl.nasa.gov
	}
\end{center}
\end{table}

\newpage
\begin{figure}
   \centering
    \includegraphics[width=3.in]{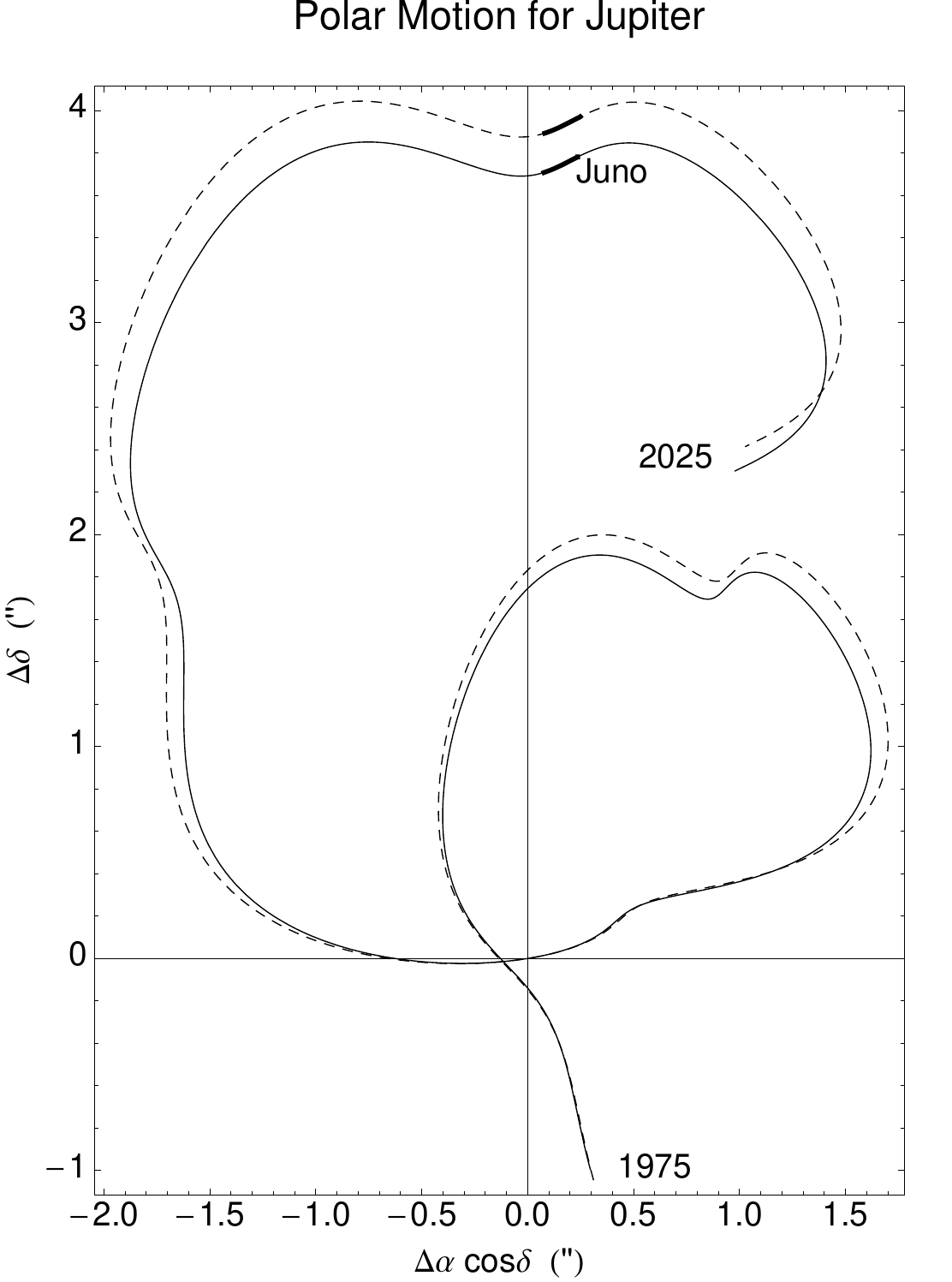}
    \caption[sat]{Nonlinear IAU Motion of the pole of Jupiter with torques from the Sun and Galilean satellites. $\alpha$ and $\delta$ are Jupiter's right ascension and declination, respectively. The predicted pole location from 1975 to 2025 is plotted for a NMoI=0.25 (solid line) and for a NMoI decreased by 5\% (dashed line). Motion during the Juno orbiter phase is indicated by a thick solid line. The figure is derived using the IAU predicted Jupiter pole location with respect to the pole location at epoch J2000.0 (Archinal et al., 2011).}
\end{figure}

\begin{figure}[h!]
   \centering
    \includegraphics[width=3.5in]{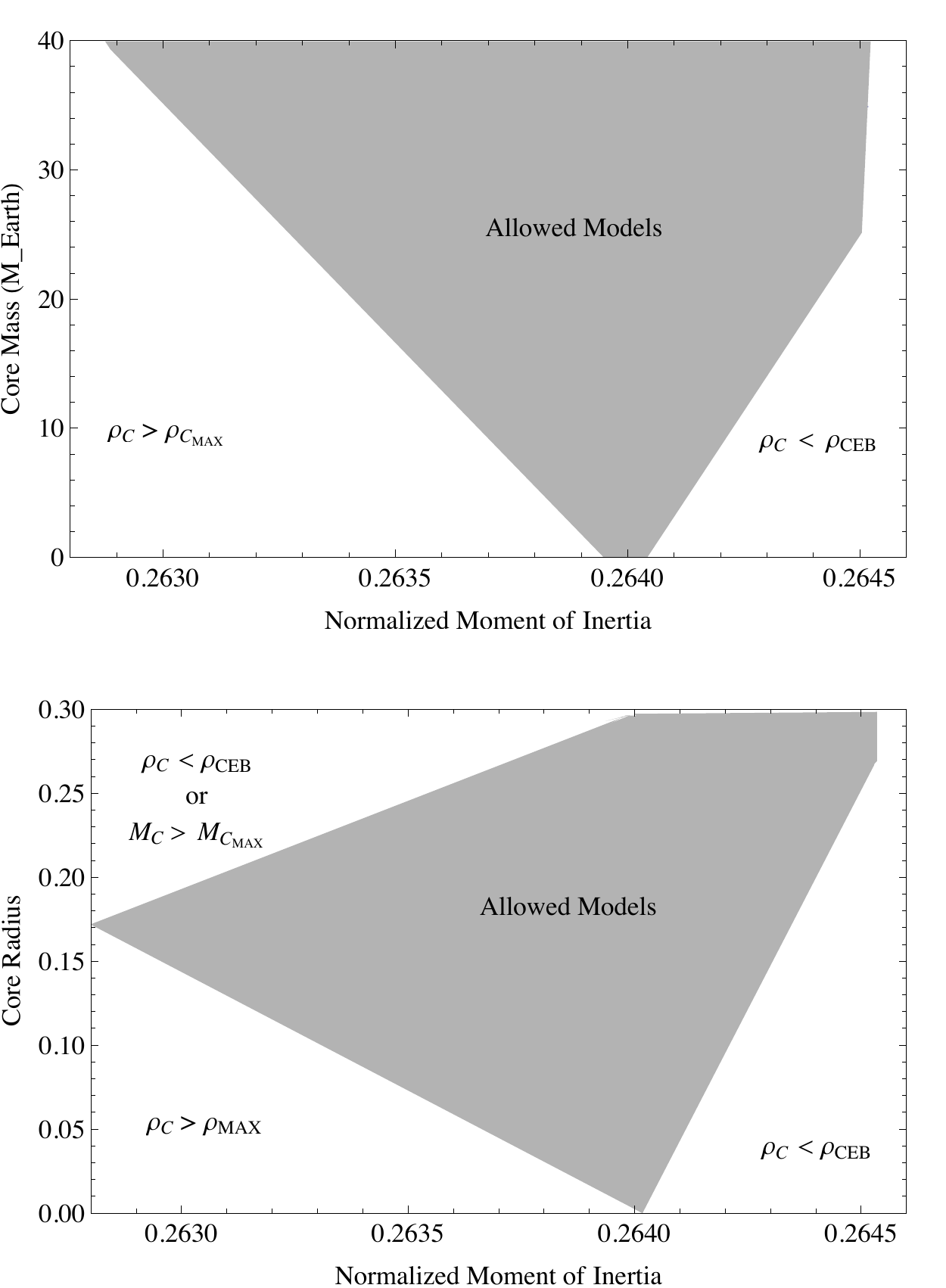}
    \caption[err]{Top: Jupiter's core mass (M$_{\oplus}$) vs. NMoI value. Bottom: Core radius (normalized to $R$) vs. NMoI. The grey region presents the range of valid models. All the allowed models fit to Jupiter's gravitational field, mass, and shape, and in addition have core properties within the allowed bounds (i.e., core mass $<$ 40 M$_{\oplus}$; core density $<$ 30 g cm$^{-3}$).}
\end{figure}

\begin{figure}[h!]
   \centering
    \includegraphics[width=2.9in]{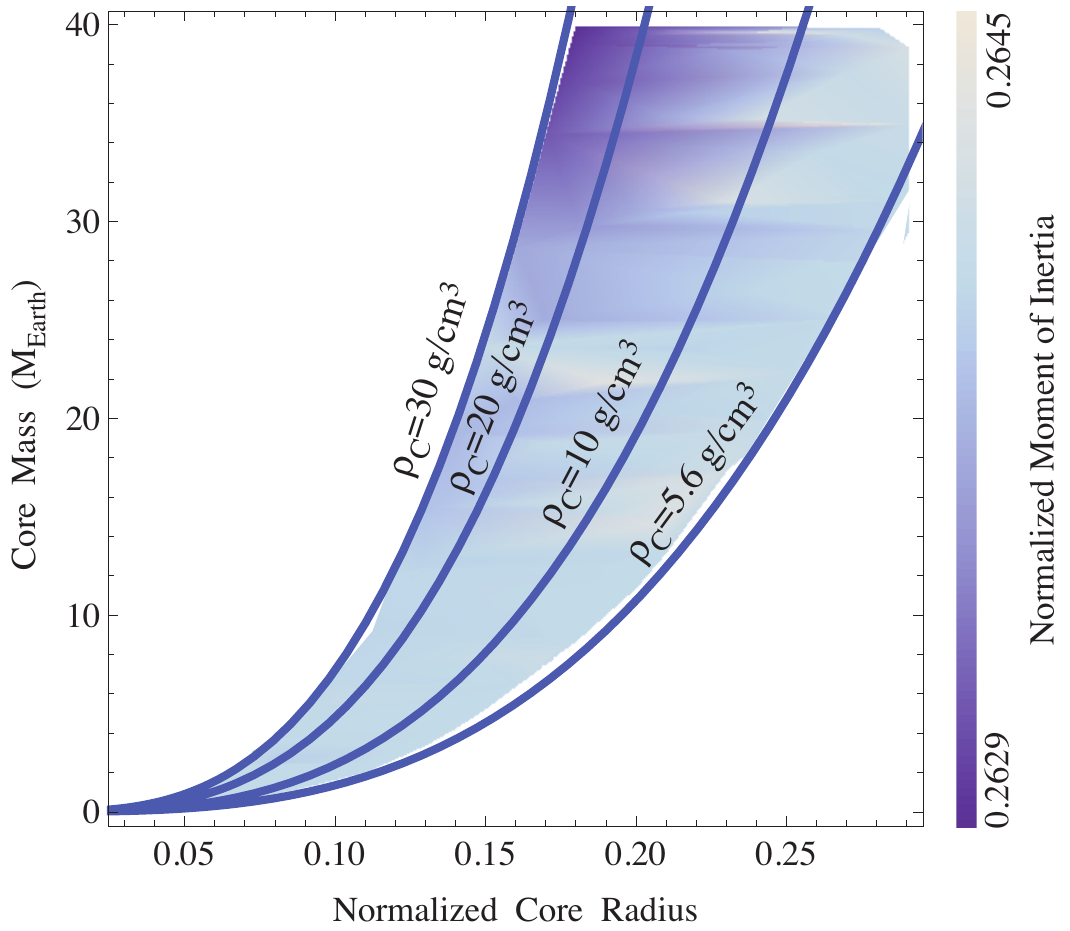}
    \caption[err]{Jupiter's calculated normalized moment of inertia in the core mass - core radius parameter space. The solid lines represent constant core densities of different values (see text for details).}
\end{figure}

  \begin{figure}[h!]
   \centering
    \includegraphics[width=3.5in]{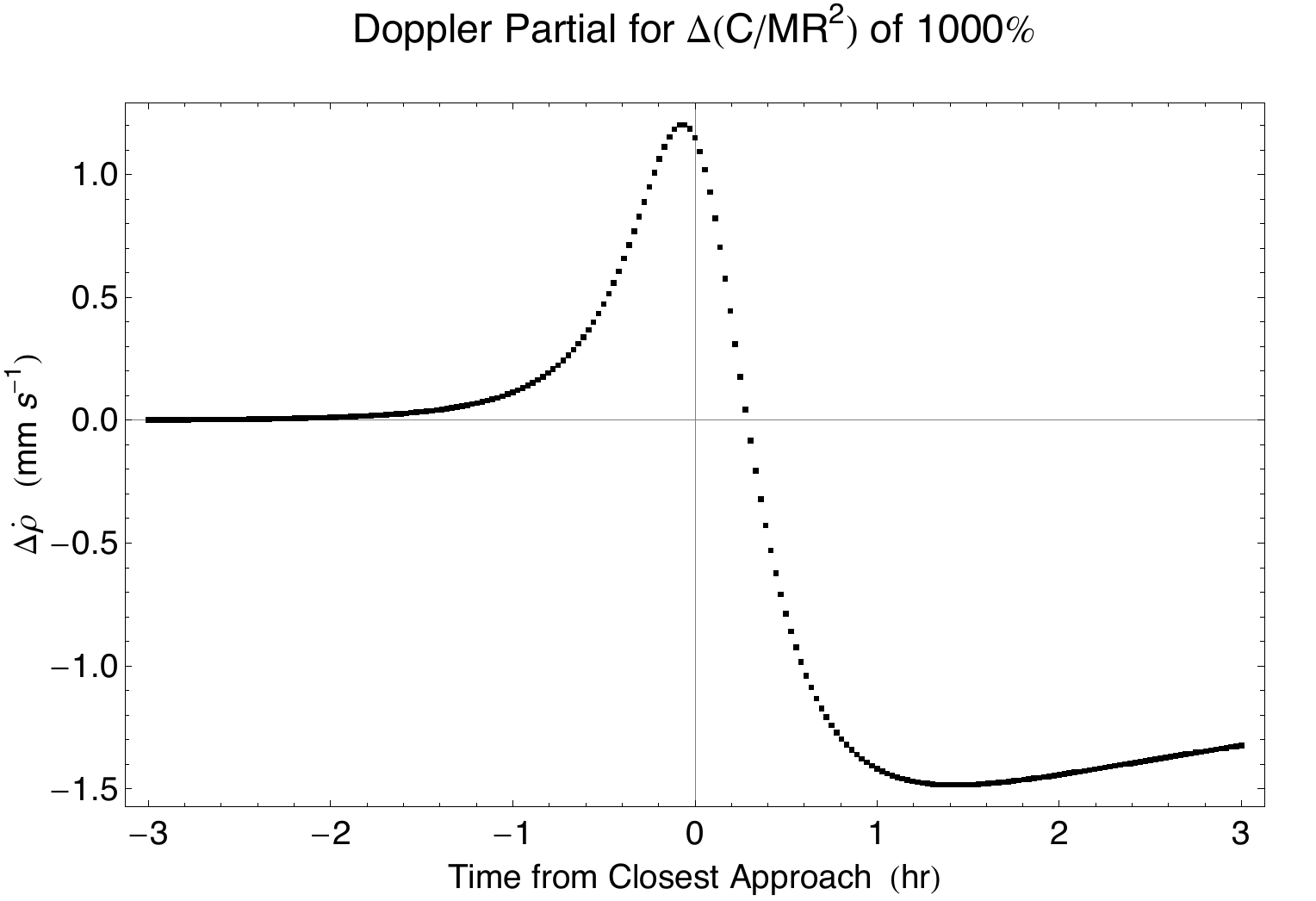}
    \caption[err]{Simulated Juno Doppler signal for anticipated perijove pass on 21 November 2016. The range rate $\rm \dot{\rho}$ is the projection of the numerically integrated spacecraft velocity on the Earth-Jupiter line of sight at perijove. }
\end{figure}
  
\begin{figure}[h!]
   \centering
    \includegraphics[width=3.5in]{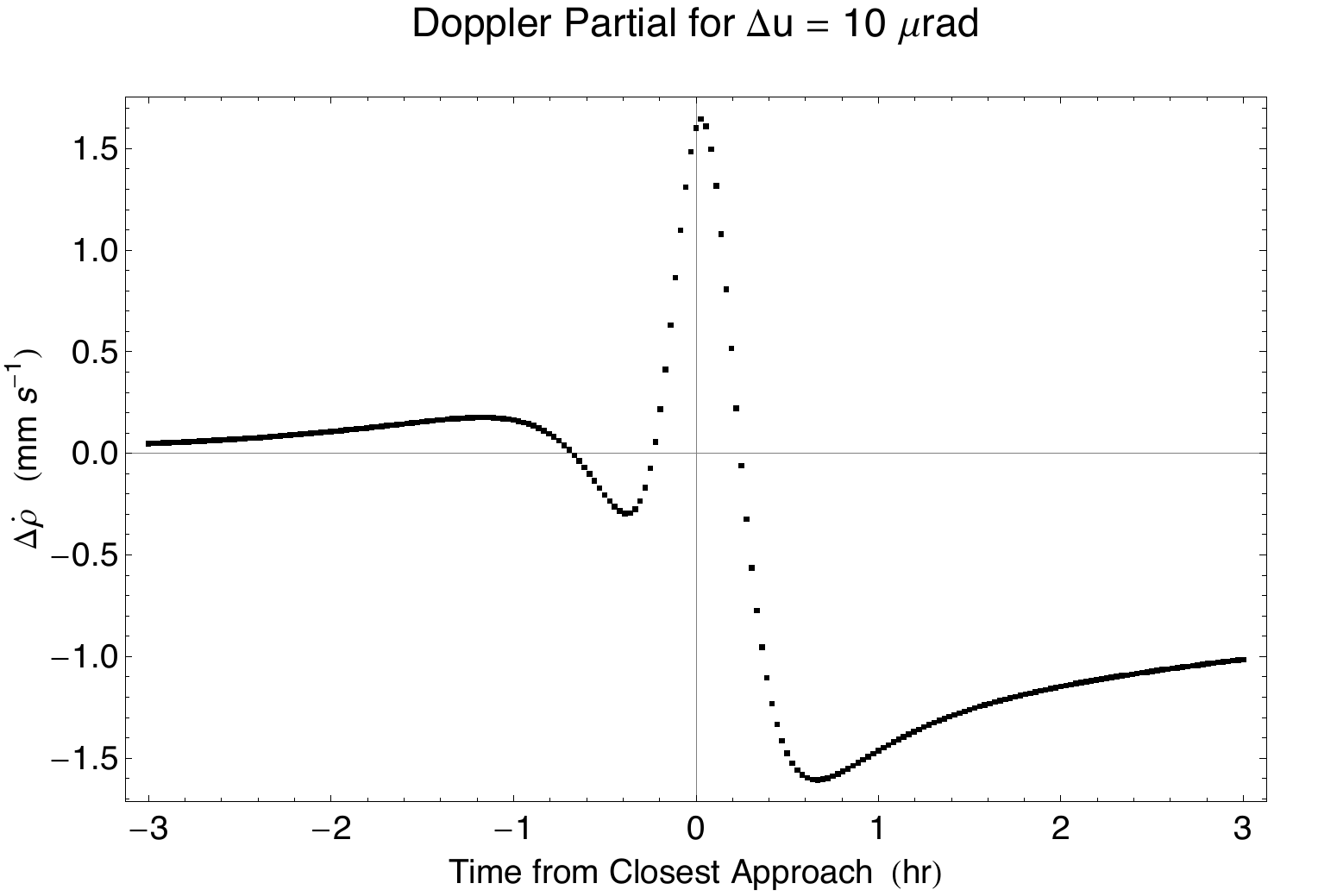}
    \caption[err]{Perturbation to the simulated Doppler signal of Fig.~1 for a change of 10 $\mu$rad in the initial pole location with respect to the IAU reference pole on 13 April 2017.}
\end{figure}

\begin{figure}[h!]
   \centering
    \includegraphics[width=3.5in]{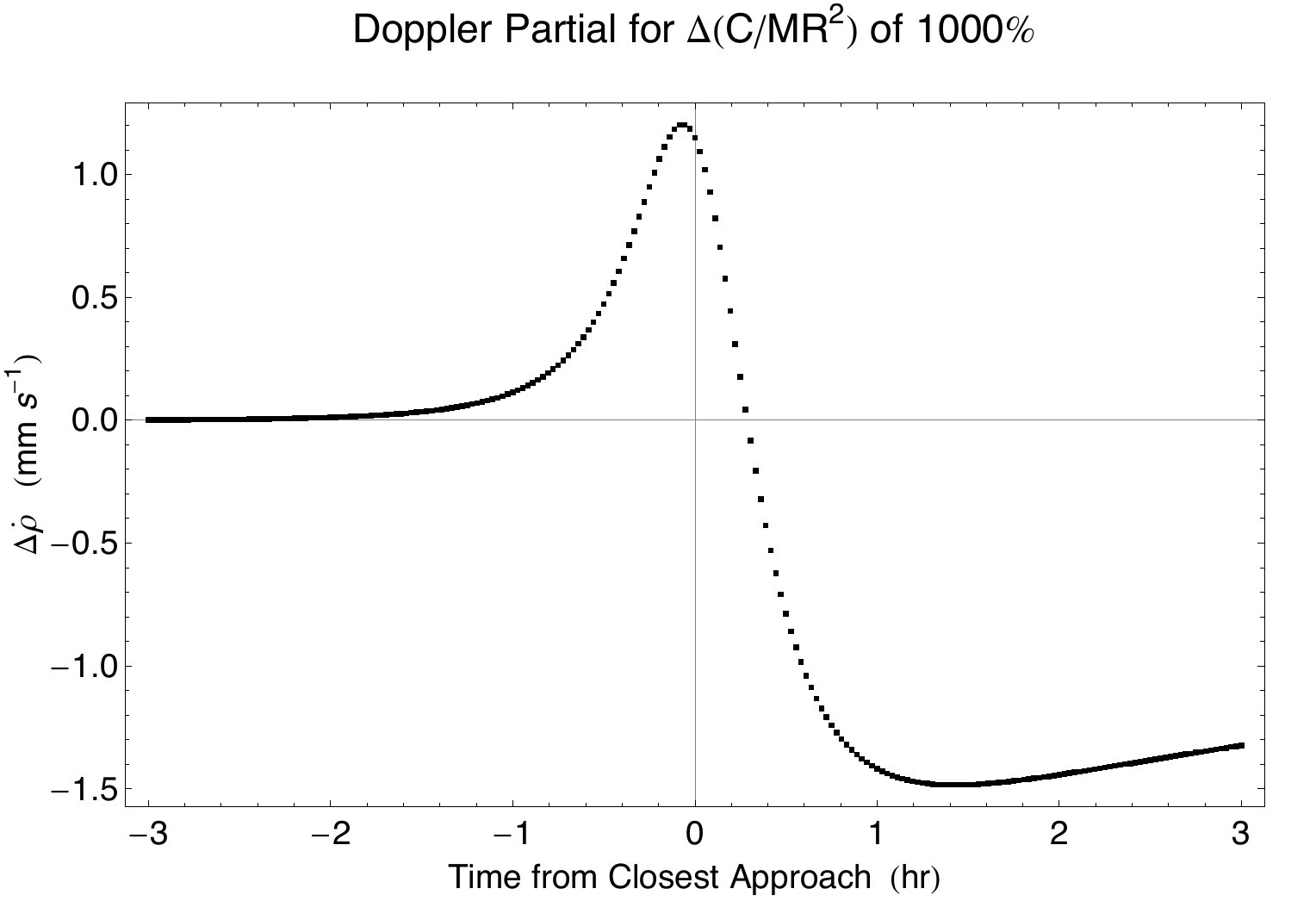}
    \caption[err]{Perturbation to the simulated Doppler signal of Fig.~1 for a change of 1000\% in the NMoI, assumed equal to 0.24.}
\end{figure}

    \begin{figure}[h!]
   \centering
    \includegraphics[width=3.5in]{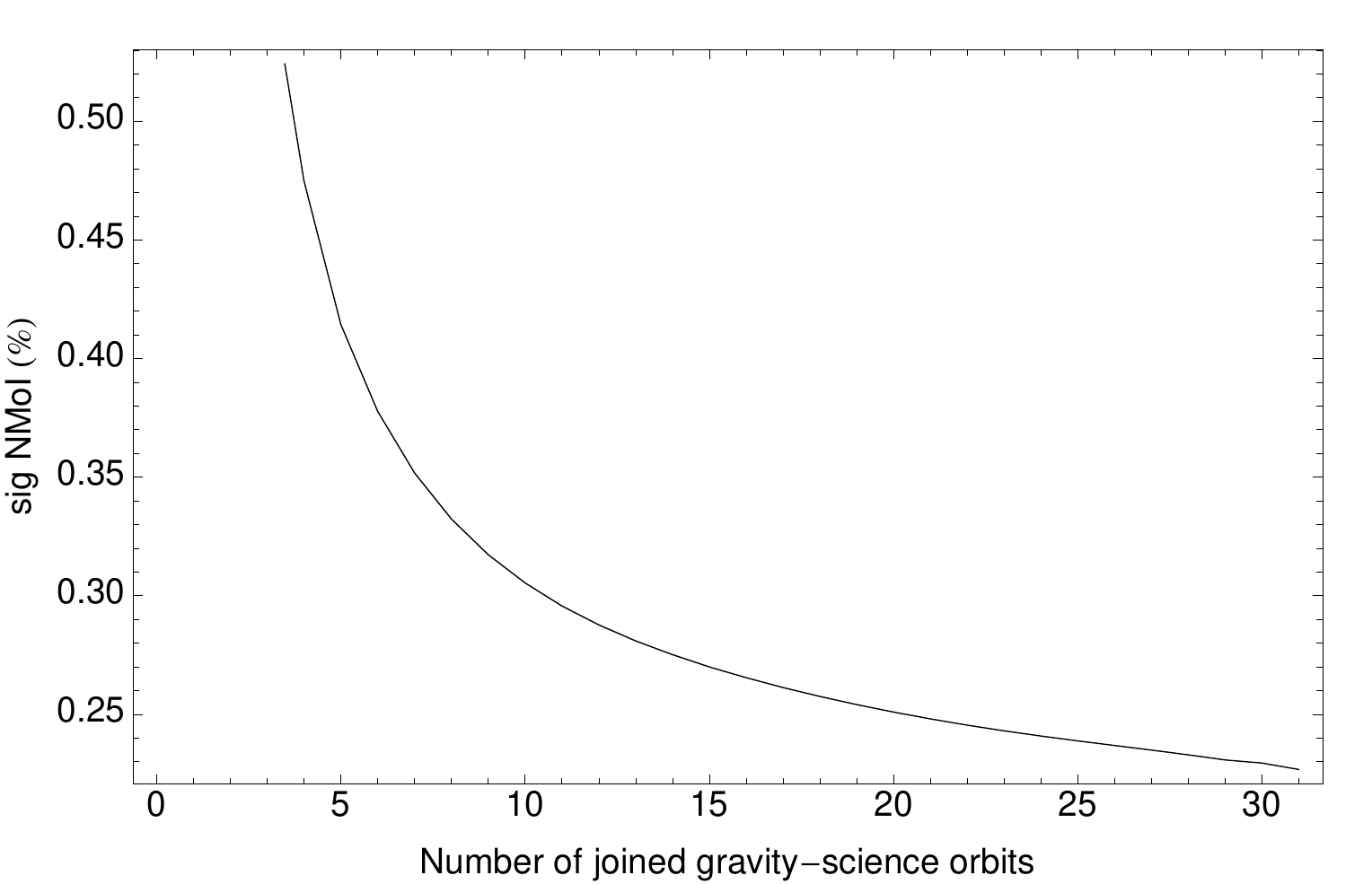}
    \caption[err]{Standard Error for Jupiter's NMoI in percents as number of gravity-science orbits joined. It is shown that a combination of the 31 orbits would reduce the standard error in NMoI by more than a factor of two.}
\end{figure}

\end{document}